%% file: main.tex
\documentclass{style/svproc}

\usepackage{cite}
\usepackage{amsmath,amssymb,amsfonts}
\usepackage{algorithmic}
\usepackage{graphicx}
\usepackage{hyperref}
\usepackage{textcomp}
\usepackage{xcolor}
\usepackage{float}
\usepackage{caption}
\usepackage{subcaption}
\usepackage{lineno}
\usepackage{booktabs}
\usepackage{enumitem}
\usepackage{multirow}
\usepackage{hyperref}
\usepackage{makecell}

\makeatletter
\newcommand{\printfnsymbol}[1]{%
  \textsuperscript{\@fnsymbol{#1}}%
}
\makeatother

\usepackage{array}
\newcolumntype{L}[1]{>{\raggedright\let\newline\\\arraybackslash\hspace{0pt}}m{#1}}
\newcolumntype{C}[1]{>{\centering\let\newline\\\arraybackslash\hspace{0pt}}m{#1}}
\newcolumntype{R}[1]{>{\raggedleft\let\newline\\\arraybackslash\hspace{0pt}}m{#1}}

\begin{document}

\input{title_page.tex}

\input{content/1_intro}

\input{content/2_related}

\input{content/3_methods}

\input{content/4_experiments}

\input{content/5_conclusion}

\bibliographystyle{style/bibtex/splncs03}
\bibliography{main.bib}

\appendix

\input{content/6_appendix}

\end{document}

%% file: title_page.tex
\title{Exploring Structure-Wise Uncertainty for 3D Medical Image Segmentation}

\titlerunning{Structure-Wise Uncertainty in Medical Imaging}

\author{
    Anton Vasiliuk \inst{1, 2} \and
    Daria Frolova \inst{1, 3} \and
    Mikhail Belyaev \inst{1, 3} \and
    Boris Shirokikh \inst{1, 3}
}


\authorrunning{A. Vasiliuk et al}

\institute{
    Artificial Intelligence Research Institute (AIRI), Moscow, Russia
    \and
    Moscow Institute of Physics and Technology, Moscow, Russia
    \and
    Skolkovo Institute of Science and Technology, Moscow, Russia
    \\
    \email{boris.shirokikh@skoltech.ru}
}

\maketitle


\begin{abstract}

When applying a Deep Learning model to medical images, it is crucial to estimate the model uncertainty. Voxel-wise uncertainty is a useful visual marker for human experts and could be used to improve the model’s voxel-wise output, such as segmentation. Moreover, uncertainty provides a solid foundation for out-of-distribution (OOD) detection, improving the model performance on the image-wise level. However, one of the frequent tasks in medical imaging is the segmentation of distinct, local structures such as tumors or lesions. Here, the structure-wise uncertainty allows more precise operations than image-wise and more semantic-aware than voxel-wise. The way to produce uncertainty for individual structures remains poorly explored. We propose a framework to measure the structure-wise uncertainty and evaluate the impact of OOD data on the model performance. Thus, we identify the best UE method to improve the segmentation quality. The proposed framework is tested on three datasets with the tumor segmentation task: LIDC-IDRI, LiTS, and a private one with multiple brain metastases cases.

\end{abstract}

\keywords{
    Uncertainty Estimation, Out-of-Distribution Detection, Segmentation, CT, MRI
}

%% file: content/1_intro.tex
\section{Introduction}
\label{sec:intro}


Advances in Deep Learning (DL) allow solving a medical image segmentation task with near human-level quality \cite{lee2017deep}. But predictions of DL models in medical imaging could not be taken blindly and assumed to be accurate. Ideally, the model is required to provide the uncertainty estimate of its output. Estimating uncertainty maps in medical image segmentation helps to solve a wide range of problems. The uncertainties are desired for a better reception by medical experts \cite{kompa2021second}, but the direct impact is hard to measure in this case. Alternatively, one uses uncertainty on a voxel-wise level to refine the segmentation map, thus improving the model's performance \cite{iwamoto2021improving}. Uncertainty maps also could be aggregated on an image-wise level, forming a standalone out-of-distribution (OOD) detection method \cite{linmans2020efficient}.


In the case of multiple objects or \textit{structures} per image (e.g., tumors, lesions), clinical tasks also require analyzing the model's output on the structure-wise level. Such cases are common in medical, especially radiological \cite{sahiner2019deep}, imaging: a brain tumor, lung cancer, organ-at-risk, or liver tumor segmentation. However, the ways of using or reporting the uncertainty on distinctly localized multiple structures are poorly explored, rising acute questions. For example, using voxel-wise uncertainty, as in \cite{iwamoto2021improving}, one can improve the segmentation quality of detected structures but cannot filter individual false positive (FP) predicted objects. In image-wise uncertainty, as in \cite{linmans2020efficient}, we do not consider the segmentation of local structures and also rebalance FP and true positive (TP) predictions in a sub-optimal way, filtering the whole image at once.

Therefore, in this work, we study uncertainty for individual predicted structures, i.e., connected areas of the predicted segmentation mask. We assume that treating uncertainty maps in a structure-wise manner helps to remove the FP detections more effectively, thus improving the detection quality. Secondly, we assume that structure-wise uncertainty (SWU) value strongly correlates with the segmentation quality of a given structure. If the latter assumption holds true, it's possible to conduct quality control to enhance the model segmentation performance in the human-in-the-loop setup \cite{leibig2017leveraging}, where the human expert refines the most uncertain (thus, worst) predictions. We validate and experimentally confirm both assumptions.



Partially, these assumptions were tested directly or indirectly in a prior work. We detail the related studies and compare with their methodology in Sec. \ref{sec:related}. We extend these studies in several major ways and below we detail our contributions:

\let\labelitemi\labelitemii
\begin{itemize}
    
    \item \textit{Structure-wise uncertainty estimation.} We evaluate different uncertainty estimation (UE) techniques and local uncertainty aggregation functions. We show that switching from predicted values space to the structure's Entropy produces 3\% fewer FP predictions on average, up to 7\% fewer on LiTS dataset, adding a negligible overhead and being applicable to any segmentation network.
    
    \item \textit{Uncertainty under out-of-distribution.} We propose to evaluate aleatoric and epistemic performance by testing on in-distribution (ID) and OOD data. We develop three OOD aleatoric setups to demonstrate different SWU properties. We show that Pairwise-Dice Uncertainty \cite{roy2019bayesian} excels in the OOD setups, filtering out 6\% more FP predictions than the baseline method, and itself in the ID setups.

    \item \textit {Extensive and robust evaluation.} We compare state-of-the-art UE techniques on three large datasets with volumetric medical images. The datasets relate to the described problem and contain cases with multiple lesions.
    
\end{itemize}


%% file: content/2_related.tex
\section{Related work}
\label{sec:related}


One of the direct SWU applications is FP reduction. Pursuing this goal, Nair et al. \cite{nair2020multiplse_sclr} improved performance of their model in the multiple sclerosis segmentation task. The authors took a sum of logarithms (\textit{sum-log}) over a predicted structure uncertainty as a score to filter them. We argue that the sum-log is biased towards small objects. In a broader setup with the differently sized target structures, we show that the standard aggregation techniques such as \textit{mean} surpass sum-log with a great margin.

Another approach to filter FP is a dedicated postprocessing model. Ozdemir et al. \cite{ozdemir2017propagating} trained a network to classify predicted structures and compared different dropout and ensembling regimes for this network. Bhat et al. \cite{bhat2021liver_fp} reduced FP in the liver lesion segmentation task by training an SVM classifier on predicted patches, their uncertainties, and hand-crafted features. However, FP reduction with a separate network is limited with strictly one structure per patch or image. Here, we consider a more general setup.

Other studies explore the ability to predict quality from uncertainty. Roy et al. \cite{roy2019bayesian} developed a Monte-Carlo-based approach to predict whole-brain segmentation and uncertainty maps. The authors calculated mean entropy, pairwise Dice score, coefficient of the volume variation, and intersection over union to predict structure-wise Dice scores. Mehrtash et al. \cite{mehrtash2019confidence} proposed to use mean entropy to predict structure Dice scores and achieved a high Pearson correlation between them for different tasks.
Hoebel et al. \cite{hoebel2020exploration} studied several setups for the whole image quality prediction. They compared Deep Ensembles against Monte-Carlo dropout and Dice loss against weighted cross-entropy in terms of pairwise Dice score, coefficient of volume variation, and mean entropy value. DeVries et al. \cite{devries2018LeveragingUE} trained a separate network to predict image-wise segmentation quality and compared different uncertainty estimation methods with this network. We extend these approaches by studying uncertainty application in a structure-wise manner instead of the image-wise one and evaluate all related UE techniques. Moreover, we introduce studying uncertainty in the OOD setup.

SWU is also taken advantage of in other challenges. Seeböck et al. \cite{seebock2019weakly_sup} developed an anomaly detection method for retinal optical coherence tomography, but the authors pursue the other goal of developing a weakly-supervised segmentation model. Hiasa et al. \cite{hiasa2020active_l} studied muscle segmentation in an active learning setting and proposed to use mean structure-wise variance to predict the structure's Dice score. In our work, we identify the UE technique for the supervised segmentation problem.

Thereby, we conduct an extensive study of known uncertainty estimation techniques on a structure-wise level. We perform unified experiments across individual aggregation and  uncertainty estimation techniques, emphasizing the importance of studying both aleatoric and epistemic setups.

%% file: content/3_methods.tex
\section{Methods}
\label{sec:methods}
 
In this section, we propose a general framework to estimate SWU. The estimation process consists of three steps: (i) compute a voxel-wise uncertainty map, (ii) split the segmentation map to obtain the individual structures, and (iii) aggregate the uncertainty inside every structure. The SWU scores can be further used for the FP filtration and quality estimation.




\subsection{Structure definition}
\label{ssec:methods:struct}

The ground truth structure (e.g., lesion, tumor) is defined as a connected area of the annotation mask. Similarly, a predicted structure is a connected area of the predicted segmentation mask, which can be binarized with different probability thresholds. We experimentally compared different threshold values and found out that either larger (e.g., $0.75$) and smaller (e.g., $0.25$) ones give considerably worse results than the de facto standard threshold of $0.5$. We further use the probability threshold of $0.5$ to define a predicted structure and omit the comparison of thresholds for the clarity.

\subsection{Uncertainty estimation methods}
\label{ssec:methods:ue}

To obtain uncertainty maps, we use Deep Ensembles \cite{lakshminarayanan2017simple}, which are considered to be state-of-the-art for estimating uncertainty in the medical image segmentation tasks \cite{jungo2019assessing, mehrtash2019confidence}. We construct an ensemble of $T=5$ neural networks trained with different weight initializations, during the inference time, $T$ predicted probability maps $P_1, ..., P_T$ are generated for an input image. If probability map is a multi-channel (softmax) output, the different channels are denoted as $P^c_i$.

The conventional way to filter FP predictions is to threshold a predicted mask with its maximum value; thus, we consider \textbf{Pred (max)} a baseline method. We also use the output of the final layer before sigmoid activation instead of probabilities and call this method \textbf{Logit}. As one of the standard UE methods, we include \textbf{Entropy}: $U_{\text{Ent}} = -\sum_{c=1}^C{P^c\log P^c}$.

The methods above can be applied both to a single and the ensemble's (i.e., the average) prediction by substituting $P^c$ with $\bar{P^c}$, where $\bar{P^c} = \frac{1}{T}\sum_{t=1}^T{P^c_t}$. Alternatively, we can apply averaging after calculating the entropy: $ U_{\text{AE}} = -\frac{1}{T}\sum_{t=1}^T \sum_{c=1}^C{P^c_t \log P^c_t}$. We call this method \textbf{Average entropy (AE)} and also include it into consideration. 

The following two methods are drawn from the related work on UE and operate only on multiple predicted probability maps. The first is \textbf{Mutual Information (MI)} or BALD \cite{houlsby2011bayesian}: $U_{\text{MI}} = -\sum_{c=1}^C{\bar{P^c}\log \bar{P^c}} + \frac{1}{T}\sum_{t=1}^T\sum_{c=1}^C{P^c_t \log P^c_t}$. The second is \textbf{Voxel-wise variance} of predictions \cite{smith2018understanding}: $U_{\text{Var}} = \frac{1}{TC}\sum_{t=1}^T \sum_{c=1}^C{(P^c_t - \bar{P^c})^2}$


The last method that we consider is \textbf{Pairwise Dice (PD)} between predictions \cite{roy2019bayesian}. Unlike previous methods, it produces a single uncertainty score per structure instead of a voxel-wise uncertainty map. This uncertainty score is the averaged dice score between all pairs of $T$ predictions and a given structure.






\subsection{Uncertainty aggregation techniques}
\label{ssec:methods:aggregation}

Assuming uncertainty map is given, we need to assign a single score for every structure. In \cite{nair2020multiplse_sclr}, the authors calculate \textit{sum-log} of uncertainties for all voxels $v$ in the structure $S$: $u = \sum_{v \in S}{\log U_v}$. We argue that \textit{sum-log} is heavily unbalanced in cases with differently sized structures, which are common. Therefore, we include in comparison the standard and, in this case, balanced statistics: \textit{min}, \textit{max}, \textit{mean}, and \textit{median}.

%% file: content/4_experiments.tex
\section{Experiments}
\label{sec:exp}

\subsection{Data}
\label{sec:data}

We study SWU performance on three different challenges. To explore a method's aleatoric performance compared to epistemic, we provide an OOD dataset in every task. The model is trained only on the ID training set, and we compare its performance on the ID test set and the OOD data. All OOD datasets share the same preprocessing steps with their ID pairs; the preprocessing is disclosed in the supplementary materials.

\textbf{Mets} (\textit{private} ID dataset) includes $1554$ T1-weighted head MR images with annotated metastases masks. 
Besides, one may consider a recently published public alternative \cite{lu2021intracranial}. 
\textbf{EGD} (OOD for \textit{Mets}) includes $374$ images of brain MRI ($4$ different modalities) with annotated glioblastoma masks \cite{van2021erasmus}. We select $141$ of them with Flair as the primary modality. We consider it to have empty metastases masks. 
\textbf{LIDC} (ID) includes $1018$ chest CT images from LIDC/IDRI database \cite{armato2011lung} with annotated lung cancer masks. 
\textbf{MIDRC} (OOD for \textit{$LIDC$}) includes $110$ chest CT images with annotated COVID-19 lesion masks \cite{tsai2021rsna}. We select $98$ of them with non-empty segmentation masks. We consider it to have empty lung cancer masks. 
\textbf{LiTS} (ID) includes $131$ abdominal CT images with annotated liver and liver tumor masks \cite{bilic2019liver}. \textbf{LiTS-mod} (OOD for \textit{LiTS}) is a synthetically created dataset from $13$ \textit{LiTS} images with empty liver tumor masks, generating typical CT imaging artifacts\cite{saparov2021zero, pimkin2020multidomain}. 

All considered ID datasets are diverse and have the multiple small structures segmentation task, which satisfies the considered setup. Five out of six datasets are publicly available, yielding the partial reproducibility of our experiments.


\subsection{Experimental setup}
\label{sec:metrics}

In all our experiments, we use the same segmentation model based on nnU-Net \cite{isensee2021nnu}. The implementation and training details are provided in the supplementary materials and they are also available in our repository\footnote{\url{https://github.com/BorisShirokikh/u-froc}}.


\paragraph{Metrics.}

To measure FP reduction capacity, we evaluate how many FP detections per image are filtered at 95\% recall level and compute average recall for high precision values. Considering $R_{\text{max}}$ is the maximum model's recall value, and $F_x$ is the average number of FP predictions for a recall value $\frac{x}{100} \times R_{\text{max}}$, we compute the FP reduction metric as $\frac{F_{100} - F_{95}}{F_{100}}$ to account for a different number of FP on the OOD setups. The average recall is computed for precision values $P$ from $\text{min} (P)$ to $\frac{1}{2}(\text{min} (P) + \text{max} (P))$, the same for each method on a setup, to obtain statistics only from the more relevant high recall region. For quality control metrics, we report the absolute value of the Spearman correlation coefficient between the individual structure Dice scores and SWU values.

\subsection{Results}

\subsubsection{FP reduction.}

\begin{figure}[t]
    \centering
    \includegraphics[scale=.24]{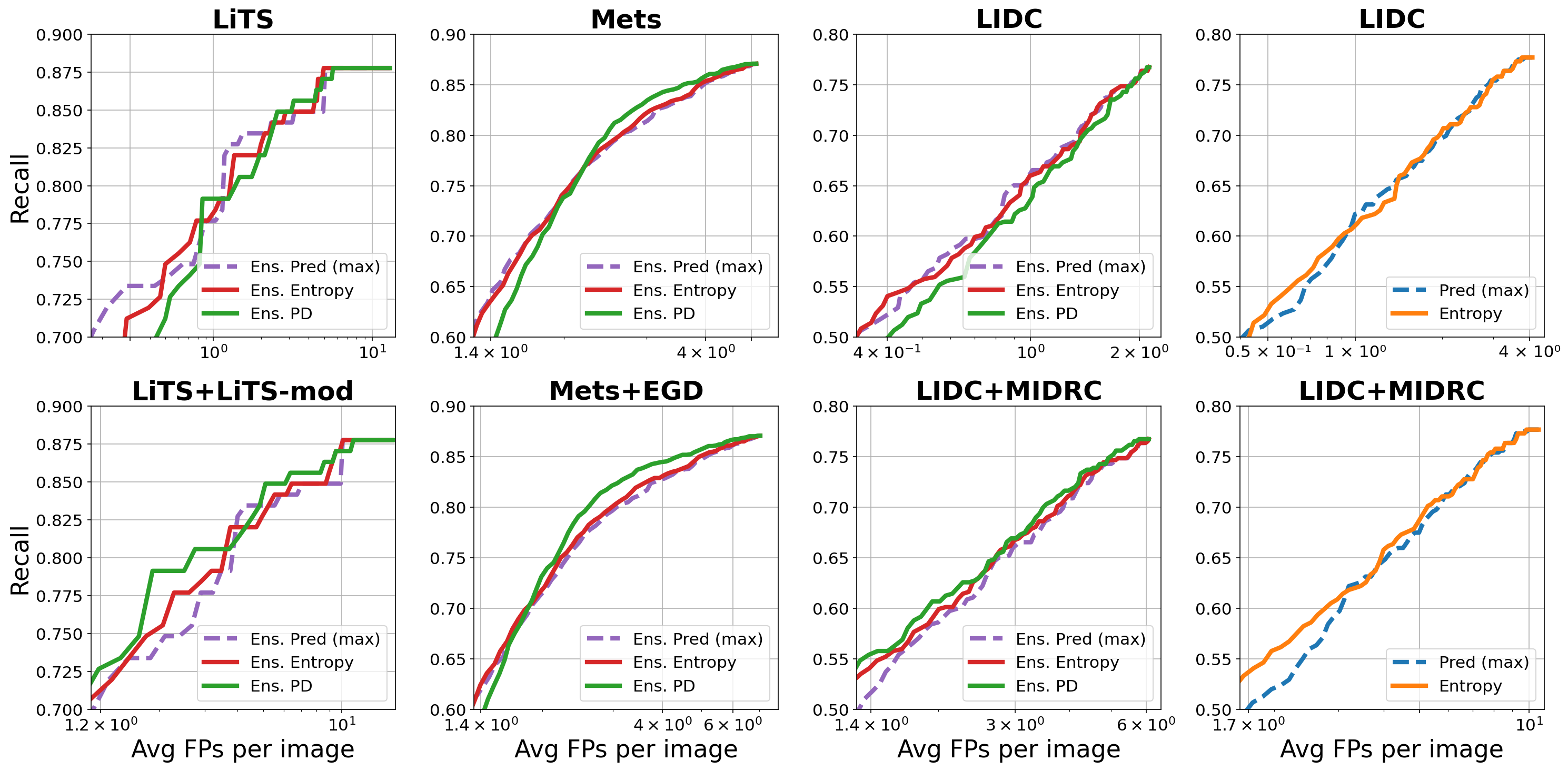}
    \caption{FROC curves for the best methods in comparison with the baseline (the dashed line). For visual clarity, the average number of FP per image is given in the log scale. The measures are obtained on the ID data (row 1) and the ID and OOD data combined (row 2). \textit{Discrepancy methods show better performance when OOD dataset is present.}
    }
    \label{fig:frocs}
\end{figure}
\newcommand{\bd}[1]{\textbf{#1}}

Despite the solid performance of the baseline method, there are advantages of using other uncertainty measures and aggregation techniques; see Tab.~\ref{table:frocs}. Using the Entropy measure or mean aggregation, one can produce fewer FP predictions for most setups. Except for Variance, AE, and \textit{Entropy (sum-log)} \cite{nair2020multiplse_sclr}, the other methods surpass the baseline. The most consistent methods are PD and Entropy, allowing for up to 7\% FP reduction with a single model and 11\% with the ensemble model.

A considerable rise in OOD performance is shown by the \textit{discrepancy} methods (PD, MI, Variance) in comparison to the \textit{averaging} methods (Pred, Logit, AE, Entropy); see Fig. \ref{fig:frocs}, Tab.~\ref{table:frocs}. The discrepancy methods produce 6-8\% fewer FP predictions on the aleatoric OOD setups while averaging methods do not exceed 3\% limit, with an even more apparent difference on individual datasets. Since the OOD setups differ from the ID ones only in the additional FP samples, we can conclude that the discrepancy methods are better at filtering OOD data.

\begin{table}
\centering
\caption{FP reduction capabilities of different SWU methods. Metrics are described in Sec.~\ref{sec:metrics}. Datasets with added OOD counterpart data are marked by *. \textit{Mean} aggregation is used unless it is stated in brackets. Average FP reduction for ID and OOD setups is provided in the Epi. and Alea. columns, respectively.}

\resizebox{\textwidth}{!}{%
\begin{tabular}{lc|cc|cccccc|cccccc}
\toprule
{Method} &   & \multicolumn{ 8}{c}{ FP reduction }  &   \multicolumn{ 6}{c}{Average Recall} \\

{} & {Ensemble} & \textbf{Epi.} & \textbf{Alea.} & LiTS &  LiTS* &  LIDC&  LIDC* &  Met&  Met* &   LiTS &  LiTS* &  LIDC&  LIDC* &  Mets&  Mets* \\
\midrule
Pred (max) & & \large{.52} & \large{.54} & .75 &  .78 & .36 &   \textbf{.33} &  .45 &   .51 &                        \bd{.86} & \bd{.84} & .69 &  .58 &  .83 & .80 \\
Pred   &   & \large{.53} & \large{.56} & .80 &    .82 &  .31 &  .30 & \textbf{.48} &   \textbf{.56}  &                 \bd{.86} & \bd{.84} & .69 &  .61 & \bd{.84} & \bd{.81} \\
Logit  &   & \large{.53} & \large{\textbf{.57}} &  .79 &    .82 &  .31 &  .30 & \textbf{.48} &   \textbf{.56} &                   \bd{.86} & \bd{.84} & .69 &  .61 & \bd{.84} & \bd{.81} \\
Entropy &   &  \large{\textbf{.54}} & \large{\textbf{.57}} & \textbf{.82} &   \textbf{.83} &  .32 &  .32 &  \textbf{.48} &   \textbf{.56} & \bd{.86} & \bd{.84} & .69 & \bd{.62} & \bd{.84} & \bd{.81} \\
Entropy (min) &    & \large{.52} & \large{.54} &  .77 & .79 &  .36 &  \textbf{.33} &  .44 &   .51 & \bd{.86} & \bd{.84} & .69 & .59 & .83 & .80 \\
Entropy (sumlog\cite{nair2020multiplse_sclr})  & &  \large{.51} & \large{.51}  &   .75 &    .75 &  \textbf{.37} &  \textbf{.33} & .41 &   .45  &         .85 &   .83 & \bd{.70} &  .47 &  .81 & .71 \\
\midrule
Pred (max) & \checkmark & \large{.49}  & \large{.50}  &  .83 &    .81 &   .27 &   .26 &  .37 &      .44 &     .86 &      .85 &     .67 &    .58 &    .83 &  .82 \\
Pred     & \checkmark &  \large{.49}  & \large{.52} & .82 &    .81 &   .27 & .29 &  .39 & .47 & \bd{.87} &      .85 &     .67 &    .60 &    .83 &  .83 \\
Logit    & \checkmark &  \large{.50} & \large{.53} & .83 &    .82 &   .27 & .29 &  .39 & .47 & \bd{.87} &      .85 &     .67 &    .61 & \bd{.84} &  .83 \\
AE (min)  & \checkmark & \large{.44} & \large{.41} &.85  &   .83 &   .22 & .13 &  .26 &  .28 &   .86 &   .84 &   .67 &    .55 &    .82 &  .81 \\
Entropy  & \checkmark &  \large{.50} & \large{.53} & .84 &    .82 &   \textbf{.28} & .29 &  .39 &  .47 &  \bd{.87} &  .85 &     .67 &    .61 &  .83 & \bd{.84} \\
Entropy (min)  & \checkmark & \large{\textbf{.51}} & \large{.52} & \textbf{.89} & \textbf{.85} &  .26 & .27 &  .37 &  .44 &  .86 &  .85 & \textbf{.68} &    .58 &  .82 & \bd{.84} \\
Entropy (sumlog\cite{nair2020multiplse_sclr}) & \checkmark &  \large{.43} & \large{.45} & .71 &  .70 &   .27 &  .26 &  .32 &    .38 &   .86 &   .79 & \bd{.68} &   .44 &    .82 &  .79 \\
MI      & \checkmark & \large{.47} & \large{.55} &  .79 &    .79 &   .21 &  \textbf{.32} &  .42 &  .54 &  .86 &   .84 &   .64 & \bd{.63} & \bd{.84} & \bd{.84} \\
PD & \checkmark & \large{.50} & \large{\textbf{.56}} & .82 &  \textbf{.83} & .23 & .30 & \textbf{.44} & \textbf{.55} & \bd{.87} & \bd{.86} & .66 & \bd{.63} & \bd{.84} & \bd{.84}\\
Variance (min) & \checkmark &  \large{.40} & \large{.48} & .65 &    .68 &   .14 & .26 &  .41 &    .50 &   .83 &      .80 &     .63 &    .58 & \bd{.84} &  .83 \\
\bottomrule

\end{tabular}}
\label{table:frocs}
\end{table}

\subsubsection{Quality control.}

\begin{figure}[t]
    \centering
    \includegraphics[scale=.22]{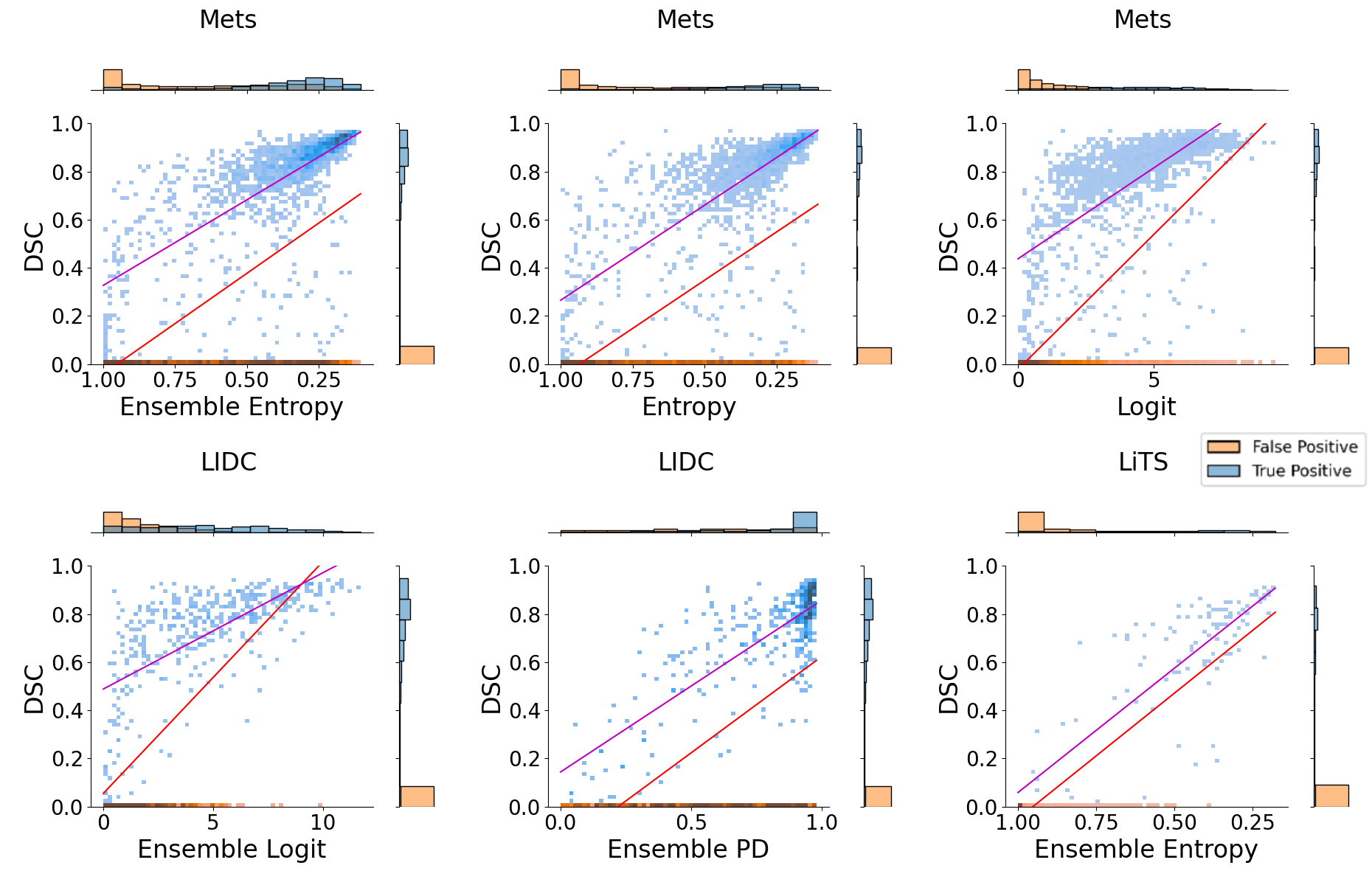}
    \caption{Linear models for object Dice scores from SWU plotted on top of the structures' heat-maps. The purple and red lines are constructed for only TP and both TP and FP structures, respectively. Single-dimensional distributions of the measures are plotted along the axes.}
    \label{fig:corrs}
\end{figure}

For most of the methods, \textit{mean} aggregation is a better index of a structure quality than \textit{min} and \textit{max} aggregations (Fig. \ref{fig:corrs} and Tab. \ref{table:corr}). The only exceptions are Variance, with poor performance in all setups, and Pairwise Dice, which does not use the voxel space uncertainty. The \textit{Entropy (mean)} is the best method in all LIDC and Mets setups and the second best in ensemble LiTS setup, while discrepancy methods generally show the weaker correlation.

Note that we do not consider the OOD setups in this quality control study, since OOD data only introduces FP instances and, thus, does not affect the correlation of scores on TP instances.

Overall, the most consistent method to evaluate SWU is the mean Entropy. It performs among the top methods, producing 2.5\% fewer FP predictions on average and giving a $.77$ Spearman correlation with the object Dice score for TP predictions. In the presence of OOD data, Pairwise Dice score reduces FP predictions better than others, filtering from 2\% to 11\% more FP structures, depending on the OOD setup.

\begin{table}
\centering
\caption{Spearman correlation coefficients between structure's Dice score and SWU value for TP predictions. Columns denoted by ``**'' show values for all predictions, including FP. The values separated by ``/'' represent \textit{mean} and extreme aggregation, respectively.}
\resizebox{0.9\textwidth}{!}{%
\begin{tabular}{lcc|@{\hskip .1in}c@{\hskip .1in}c@{\hskip .1in}c@{\hskip .1in}c@{\hskip .1in}c@{\hskip .1in}c @{\hskip .1in}}
\toprule
{} & Second Agg. & Ensemble & LiTS &  LiTS** &  LIDC &  LIDC** &  Mets &  Mets** \\
\midrule
Pred  & max &  &  
\textbf{.86}/\scriptsize{.81} & .46/\scriptsize{.46} & .68/\scriptsize{.62} & .63/\scriptsize{.63} & \textbf{.79}/\scriptsize{.65} & .61/\scriptsize{.60}  \\
Logit & max & &  
 .85/\scriptsize{.81} & .46/\scriptsize{.46} & .67/\scriptsize{.62} & \textbf{.64}/\scriptsize{.63} & .75/\scriptsize{.65} & .61/\scriptsize{.60}  \\
Entropy & min &  &  
\textbf{.86}/\scriptsize{.81} & .46/\scriptsize{.46} & \textbf{.69}/\scriptsize{.62} &\textbf{.64}/\scriptsize{.63} & \textbf{.79}/\scriptsize{.65} & .61/\scriptsize{.60}  \\
\midrule
Pred  & max & \checkmark   &
\textbf{.79}/\scriptsize{.69} & .66/\scriptsize{.66} & \textbf{.70}/\scriptsize{.64} & .66/\scriptsize{.66} & \textbf{.78}/\scriptsize{.64} &\textbf{.61}/\scriptsize{.59}  \\
Logit  & max & \checkmark &  
 .75/\scriptsize{.69} & .66/\scriptsize{.66} & .69/\scriptsize{.64} & \textbf{.67}/\scriptsize{.66} & .74/\scriptsize{.64} & \textbf{.61}/\scriptsize{.59} \\

AE & min & \checkmark &  
 .72/\scriptsize{.69} & .55/\scriptsize{.65} & .55/\scriptsize{.61} & .58/\scriptsize{.65} & .77/\scriptsize{.64} & .53/\scriptsize{.58} \\

Entropy & min & \checkmark &  
.78/\scriptsize{.69} & .66/\scriptsize{.66} & \textbf{.70}/\scriptsize{.64} &\textbf{.67}/\scriptsize{.66} & \textbf{.78}/\scriptsize{.64} &\textbf{.61}/\scriptsize{.59} \\
MI & min & \checkmark &  
 .72/\scriptsize{.24} & .64/\scriptsize{.48} & .49/\scriptsize{.38} & .48/\scriptsize{.40} & .57/\scriptsize{.27} & .50/\scriptsize{.33}  \\
PD & min & \checkmark  &   
 .76/\scriptsize{.74} & .65/\scriptsize{\textbf{.71}} & .67/\scriptsize{.66} & .64/\scriptsize{.63} & .74/\scriptsize{.76} & .59/\scriptsize{.60}   \\
Variance & min & \checkmark &  
 .42/\scriptsize{.64} & .44/\scriptsize{.60} & .22/\scriptsize{.55} & .22/\scriptsize{.55} & .10/\scriptsize{.52} & .22/\scriptsize{.54}  \\
\bottomrule
\end{tabular}}
\label{table:corr}
\end{table}

\subsection{Discussion}
\label{sec:discussion}

To construct OOD setups with positive samples, we had to include the ID data. That means that the FP reduction metric shows an average between ID and OOD false positives, and pure OOD performance remains unknown. One of the possible ways to approach this problem is to create a domain-shifted setup which would contain OOD data with the true-positive structures.

The other promising application of the SWU framework is a more efficient human-in-the-loop control. Quality estimates might be a good measure to select images or individual structures to show a medical professional, but the question of how to gain the most quality given a limited amount of human interaction, combined with optimal FP reduction remains open.




%% file: content/5_conclusion.tex
\section{Conclusion}
\label{sec:conclusion}

In this work, we have conducted an extensive study of structure-wise uncertainty over six different setups. We have shown that mean Entropy provides a solid baseline in both false positive reduction and quality control tasks. Also, we have revealed the importance of studying uncertainty metrics under different origins of data. In our experiments, the discrepancy SWU methods perform significantly better for FP reduction in the presence of the OOD data, with the best results achieved by Pairwise Dice. Provided results should serve as a solid baseline for future structure-based analysis.



\paragraph{Acknowledgements.}

The authors acknowledge the National Cancer Institute and the Foundation for the National Institutes of Health, and their critical role in the creation of the free publicly available LIDC/IDRI Database used in this study.
This research was funded by Russian Science Foundation grant number 20-71-10134.


%% file: content/6_appendix.tex
\section{Experimental setup}

\subsection{Preprocessing}
Here, we describe data preparation steps including datasets splits, normalization, and interpolation.

\textbf{Mets} data is randomly split into train ($1140$ images) and test ($414$ images) sets. We interpolate the images to have $1 \times 1 \times 1$ mm spacing.

\textbf{LIDC} data is randomly split into train ($812$ images) and test ($204$ images) sets. We clip image intensities between $-1350$ and $350$ Hounsfield units (HU) -- the standard lung window. We interpolate images to have $1 \times 1 \times 1.5$ mm spacing.

\textbf{LiTS} is presented as two subsets, so we use the first as a test ($28$ images) and the second, excluding cases with empty tumor masks, as a train ($90$ images) set. The images are cropped to the provided liver masks. The intensities are clipped to the $[-150, 250]$ HU interval -- the standard liver window. Finally, we interpolate images to have $0.77 \times 0.77 \times 1$ mm spacing.

\textbf{LiTS-mod} is obtained by random changes of the reconstruction kernel to be extremely soft ($a=-0.7, b=0.5$) or sharp ($a=30, b=3$) using the implementation and notations of \cite{saparov2021zero}, and addition of ``metal'' artifacts (ball of radius $5$ and $3000$ HU) by substituting the parts of sinogram projection, as in \cite{pimkin2020multidomain}. 

Before passing through the network, we scale image intensities in $[0, 1]$.

\subsection{Training setup}
Although using cross-entropy loss has theoretical justifications of encouraging better calibrated predictions \cite{lakshminarayanan2017simple}, models trained with this loss function fail in our segmentation task. For that reason we use Dice Loss \cite{milletari2016v} and its modifications in our experiments. Thus, uncertainty estimates might be shifted in such tasks, and experimental evaluation, as in our study, becomes even more relevant. 
All models are trained in a patch-based manner: patches are sampled randomly so that they contain structures. We use SGD optimizer with Nesterov momentum of $.9$ and $10^{-3}$ initial learning rate, which is decreased to $10^{-4}$ after $80\%$ of epochs. For LiTS and Mets segmentation the model is trained for $100$ epochs ($100$ iterations per epoch, batch size $20$), while for LIDC segmentation there are $30$ epochs ($1000$ iterations per epoch, batch size $2$).

%% file: main.bbl
\begin{thebibliography}{10}
\providecommand{\url}[1]{\texttt{#1}}
\providecommand{\urlprefix}{URL }

\bibitem{armato2011lung}
Armato~III, S.G., McLennan, G., Bidaut, L., McNitt-Gray, M.F., Meyer, C.R.,
  Reeves, A.P., Zhao, B., Aberle, D.R., Henschke, C.I., Hoffman, E.A., et~al.:
  The lung image database consortium (lidc) and image database resource
  initiative (idri): a completed reference database of lung nodules on ct
  scans. Medical physics  38(2),  915--931 (2011)

\bibitem{bhat2021liver_fp}
Bhat, I., Kuijf, H.J., Cheplygina, V., Pluim, J.P.: Using uncertainty
  estimation to reduce false positives in liver lesion detection. In: 2021 IEEE
  18th International Symposium on Biomedical Imaging (ISBI). pp. 663--667
  (2021)

\bibitem{bilic2019liver}
Bilic, P., Christ, P.F., Vorontsov, E., Chlebus, G., Chen, H., Dou, Q., Fu,
  C.W., Han, X., Heng, P.A., Hesser, J., et~al.: The liver tumor segmentation
  benchmark (lits). arXiv preprint arXiv:1901.04056  (2019)

\bibitem{devries2018LeveragingUE}
Devries, T., Taylor, G.W.: Leveraging uncertainty estimates for predicting
  segmentation quality. ArXiv  abs/1807.00502 (2018)

\bibitem{hiasa2020active_l}
Hiasa, Y., Otake, Y., Takao, M., Ogawa, T., Sugano, N., Sato, Y.: Automated
  muscle segmentation from clinical ct using bayesian u-net for personalized
  musculoskeletal modeling. IEEE Transactions on Medical Imaging  39(4),
  1030--1040 (2020)

\bibitem{hoebel2020exploration}
Hoebel, K., Andrearczyk, V., Beers, A., Patel, J., Chang, K., Depeursinge, A.,
  Müller, H., Kalpathy-Cramer, J.: {An exploration of uncertainty information
  for segmentation quality assessment}. In: Išgum, I., Landman, B.A. (eds.)
  Medical Imaging 2020: Image Processing. vol. 11313, pp. 381 -- 390.
  International Society for Optics and Photonics, SPIE (2020),
  \url{https://doi.org/10.1117/12.2548722}

\bibitem{houlsby2011bayesian}
Houlsby, N., Husz{\'a}r, F., Ghahramani, Z., Lengyel, M.: Bayesian active
  learning for classification and preference learning. arXiv preprint
  arXiv:1112.5745  (2011)

\bibitem{isensee2021nnu}
Isensee, F., Jaeger, P.F., Kohl, S.A., Petersen, J., Maier-Hein, K.H.: nnu-net:
  a self-configuring method for deep learning-based biomedical image
  segmentation. Nature methods  18(2),  203--211 (2021)

\bibitem{iwamoto2021improving}
Iwamoto, S., Raytchev, B., Tamaki, T., Kaneda, K.: Improving the reliability of
  semantic segmentation of medical images by uncertainty modeling with bayesian
  deep networks and curriculum learning. In: Uncertainty for Safe Utilization
  of Machine Learning in Medical Imaging, and Perinatal Imaging, Placental and
  Preterm Image Analysis, pp. 34--43. Springer (2021)

\bibitem{jungo2019assessing}
Jungo, A., Reyes, M.: Assessing reliability and challenges of uncertainty
  estimations for medical image segmentation. In: International Conference on
  Medical Image Computing and Computer-Assisted Intervention. pp. 48--56.
  Springer (2019)

\bibitem{kompa2021second}
Kompa, B., Snoek, J., Beam, A.L.: Second opinion needed: communicating
  uncertainty in medical machine learning. NPJ Digital Medicine  4(1),  1--6
  (2021)

\bibitem{lakshminarayanan2017simple}
Lakshminarayanan, B., Pritzel, A., Blundell, C.: Simple and scalable predictive
  uncertainty estimation using deep ensembles. Advances in neural information
  processing systems  30 (2017)

\bibitem{lee2017deep}
Lee, J.G., Jun, S., Cho, Y.W., Lee, H., Kim, G.B., Seo, J.B., Kim, N.: Deep
  learning in medical imaging: general overview. Korean journal of radiology
  18(4),  570--584 (2017)

\bibitem{leibig2017leveraging}
Leibig, C., Allken, V., Ayhan, M.S., Berens, P., Wahl, S.: Leveraging
  uncertainty information from deep neural networks for disease detection.
  Scientific reports  7(1),  1--14 (2017)

\bibitem{linmans2020efficient}
Linmans, J., van~der Laak, J., Litjens, G.: Efficient out-of-distribution
  detection in digital pathology using multi-head convolutional neural
  networks. In: MIDL. pp. 465--478 (2020)

\bibitem{lu2021intracranial}
Lu, S.L., Liao, H.C., Hsu, F.M., Liao, C.C., Lai, F., Xiao, F.: The
  intracranial tumor segmentation challenge: Contour tumors on brain mri for
  radiosurgery. NeuroImage  244,  118585 (2021)

\bibitem{mehrtash2019confidence}
Mehrtash, A., Wells, W., Tempany, C., Abolmaesumi, P., Kapur, T.: Confidence
  calibration and predictive uncertainty estimation for deep medical image
  segmentation. IEEE Transactions on Medical Imaging  PP,  1--1 (07 2020)

\bibitem{milletari2016v}
Milletari, F., Navab, N., Ahmadi, S.A.: V-net: Fully convolutional neural
  networks for volumetric medical image segmentation. In: 2016 fourth
  international conference on 3D vision (3DV). pp. 565--571. IEEE (2016)

\bibitem{nair2020multiplse_sclr}
Nair, T., Precup, D., Arnold, D.L., Arbel, T.: Exploring uncertainty measures
  in deep networks for multiple sclerosis lesion detection and segmentation.
  Medical Image Analysis  59,  101557 (2020),
  \url{https://www.sciencedirect.com/science/article/pii/S1361841519300994}

\bibitem{ozdemir2017propagating}
Ozdemir, O., Woodward, B., Berlin, A.A.: Propagating uncertainty in multi-stage
  bayesian convolutional neural networks with application to pulmonary nodule
  detection. CoRR  abs/1712.00497 (2017), \url{http://arxiv.org/abs/1712.00497}

\bibitem{pimkin2020multidomain}
Pimkin, A., Samoylenko, A., Antipina, N., Ovechkina, A., Golanov, A.,
  Dalechina, A., Belyaev, M.: Multidomain ct metal artifacts reduction using
  partial convolution based inpainting. In: 2020 International Joint Conference
  on Neural Networks (IJCNN). pp. 1--6. IEEE (2020)

\bibitem{roy2019bayesian}
Roy, A.G., Conjeti, S., Navab, N., Wachinger, C., Initiative, A.D.N., et~al.:
  Bayesian quicknat: Model uncertainty in deep whole-brain segmentation for
  structure-wise quality control. NeuroImage  195,  11--22 (2019)

\bibitem{sahiner2019deep}
Sahiner, B., Pezeshk, A., Hadjiiski, L.M., Wang, X., Drukker, K., Cha, K.H.,
  Summers, R.M., Giger, M.L.: Deep learning in medical imaging and radiation
  therapy. Medical physics  46(1),  e1--e36 (2019)

\bibitem{saparov2021zero}
Saparov, T., Kurmukov, A., Shirokikh, B., Belyaev, M.: Zero-shot domain
  adaptation in ct segmentation by filtered back projection augmentation. In:
  Deep Generative Models, and Data Augmentation, Labelling, and Imperfections,
  pp. 243--250. Springer (2021)

\bibitem{seebock2019weakly_sup}
Seeböck, P., Orlando, J., Schlegl, T., Waldstein, S., Bogunović, H., Riedl,
  S., Langs, G., Schmidt-Erfurth, U.: Exploiting epistemic uncertainty of
  anatomy segmentation for anomaly detection in retinal oct. IEEE Transactions
  on Medical Imaging  PP,  1--1 (05 2019)

\bibitem{smith2018understanding}
Smith, L., Gal, Y.: Understanding measures of uncertainty for adversarial
  example detection. arXiv preprint arXiv:1803.08533  (2018)

\bibitem{tsai2021rsna}
Tsai, E.B., Simpson, S., Lungren, M.P., Hershman, M., Roshkovan, L., Colak, E.,
  Erickson, B.J., Shih, G., Stein, A., Kalpathy-Cramer, J., et~al.: The rsna
  international covid-19 open radiology database (ricord). Radiology  299(1),
  E204--E213 (2021)

\bibitem{van2021erasmus}
van~der Voort, S.R., Incekara, F., Wijnenga, M.M., Kapsas, G., Gahrmann, R.,
  Schouten, J.W., Dubbink, H.J., Vincent, A.J., van~den Bent, M.J., French,
  P.J., et~al.: The erasmus glioma database (egd): Structural mri scans, who
  2016 subtypes, and segmentations of 774 patients with glioma. Data in brief
  37,  107191 (2021)

\end{thebibliography}
